\documentclass{ctr}

\usepackage{ctrfont}
\usepackage{natbib}

\usepackage{url}
\usepackage{amsmath}
\usepackage{amssymb}

\usepackage{graphicx}
\usepackage{caption}
\usepackage{subcaption}
\usepackage[labelsep=period]{caption}
\renewcommand{\tablename}{\sc Table}
\renewcommand{\figurename}{\sc Figure}

\title{Toward a flow-structure-based wall-modeled large-eddy simulation paradigm}
\shorttitle{Toward a flow-structure-based WMLES paradigm}
\author{A. Elnahhas, A. Lozano-Dur\'an \and P. Moin}
\shortauthor{Elnahhas, Lozano-Dur\'an \& Moin}

\begin{document}


\maketitle

\section{Motivation and objectives} 

Large-eddy simulations (LES) fundamentally rely on a large separation of scales. They allow the large, energy-containing eddies to evolve spatiotemporally according to their governing equations, while modeling the effect of the near-universal small scales of turbulence on those large eddies using subgrid-scale (SGS) models. When walls are present, the separation of scales between the eddies inhabiting the two ends of the energy cascade ceases to exist as the wall is approached. This is due to the scaling of both large- and small- eddying motions with distance from the wall \citep{Tennekes1972,Townsend1976}. Wall-resolved LES aims to capture the large, energy-containing eddies in the cascade at all wall-normal distances. The aforementioned behavior of the large- and small- eddying motions leads to the cost of wall-resolved LES to be only slightly more favorable than DNS in terms of scaling with the Reynolds number based on the streamwise extent, $L_x$, $Re_{L_x}^{13/7}$ and $Re_{L_x}^{37/14}$, respectively, making both prohibitively expensive in simulating high-$Re$ flows \citep{Choi2012}.

Wall-modeled LES (WMLES) attempts to circumvent this problem by extending the near-isotropic grid resolution in the outer region of the flow down to the wall while modeling the entire subgrid dynamics of the near-wall region using low-order models \citep{Bose2018}. This reduces the scaling of the cost of WMLES to $Re_{L_x}$, indicating its practical importance \citep{Choi2012}. The most commonly utilized wall models rely on the law of the wall or simplified RANS equations and make no use of the current vast knowledge on the structure of wall-bounded turbulent flows. For the current application of interest, namely attached external boundary layer flows, it is important to note that the first grid point away from the wall in WMLES typically lies within the logarithmic region of the flow.

\begin{figure}
    \centering
    \includegraphics[width = \textwidth]{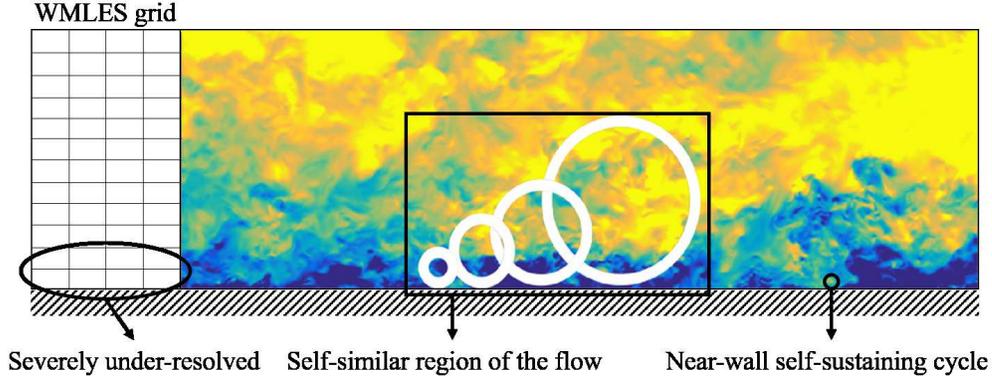}
    \caption{Instantaneous streamwise velocity in a turbulent channel flow at $Re_\tau \approx 4200$. Low velocities are indicated by darker colors and higher velocities with lighter colors. The light colored circles represent wall-attached eddies of different sizes and the dark colored circle indicates the region where the near-wall self-sustaining cycle is active. The grid on the left of the flow shows a typical uniform WMLES grid where the first grid point lies in the logarithmic region of the flow beyond the near-wall cycle and the lowest eddies in the hierarchy of attached eddies.}
    \label{fig:FlowStructure}
\end{figure}

The logarithmic region in high Reynolds number wall-bounded turbulent flows is empirically observed to be composed of different types of eddies. Most notably, the energy-containing eddies are attached to the wall and geometrically self-similar with distance from the wall \citep{Townsend1976,Marusic2019}. A synthetically generated boundary layer composed of such attached eddies allows for the prediction of the mean turbulent intensities as a function of the distance from the wall \citep{Perry1982}. Below the smallest eddies in this self-similar hierarchy, there is a near-wall self-sustaining cycle \citep{Jimenez1999}. Figure \ref{fig:FlowStructure} shows an instantaneous streamwise velocity snapshot in a turbulent channel at $Re_\tau \approx 4200$, where $Re_\tau$ is the friction Reynolds number, and depicts both the wall-attached eddies scaling with distance from the wall, as well as the region where the near-wall self-sustaining cycle is active.

\begin{figure}
    \centering
    \includegraphics[width = \textwidth]{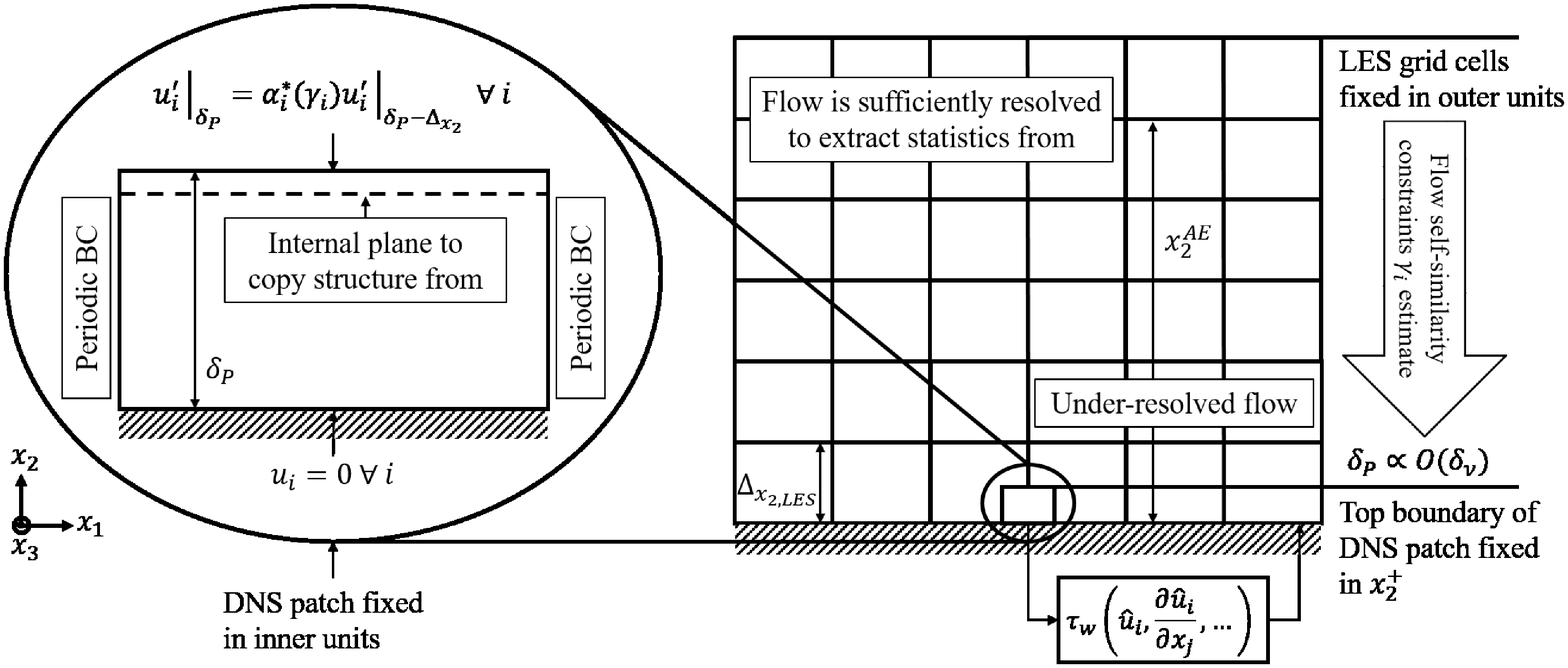}
    \caption{Schematic showing the interdependence of the DNS near-wall patch and the outer-flow LES flow field. The near-wall patch is fixed in inner units and its size is independent of the LES grid size. The outer flow centered around $x_2^{AE}$ is used to extract statistical constraints that are enforced at the top boundary of the DNS near-wall patch. The inset shows the boundary conditions of the DNS near-wall patch, where the dashed line indicates the internal plane whose structure is copied and rescaled at the top boundary.}
    \label{fig:Coupled model}
\end{figure}

Many possible reduced-order models of the near-wall cycle that make use of turbulent structures have been envisioned, but have not been used in WMLES. These models include some based on invariant solutions to the Navier-Stokes equations at low $Re_\tau$ \citep{Kawahara2011}, a statistical mean state representation of turbulence \citep{Farrell2012}, and a truncated Galerkin projection of the self-sustaining cycle \citep{Waleffe1997}. Regardless of the form that a potential near-wall cycle wall model could take, describing the near-wall dynamics using a reduced-order dynamical system model is currently limited to low-$Re_\tau$ turbulent flows or, interpreted differently, the viscous and buffer layer dynamics in high-$Re_\tau$ flows. As such, in developing a dynamical structure-based wall model that is valid for high-$Re_\tau$ flows, any near-wall cycle reduced-order model would be valid in a region that is substantially smaller than the near-wall LES grid, and would only constitute a portion of the flow that needs modeling. Furthermore, coupling such reduced-order models to the LES flow would require boundary conditions defined at the edge of the buffer layer, which would be subject to large modeling errors. Instead, if the reduced-order model contains buffer layer dynamics as well as a portion of the self-similar hierarchy of eddies, a coupling can be more easily formulated between the the reduced-order model and the LES flow, which captures the largest of these self-similar eddies away from the wall. However, since the LES flow field near the wall is under-resolved even with respect to the largest structures of the flow, a coupling between the near-wall reduced-order model and the LES flow field cannot directly depend on the flow variables computed on the LES grid point nearest to the wall. This is illustrated in Figure \ref{fig:Coupled model}.
 
We propose to use knowledge of the wall-normal self-similarity of high-$Re_\tau$ wall-bounded turbulent flows as the bridge between the LES flow field and the model representing the near-wall cycle. As a surrogate for any potential reduced-order model of the near-wall cycle, a near-wall patch of DNS resolution is used. The domain extent of this DNS patch is fixed in inner units in all three directions and is independent of the LES grid size, capturing only the smallest eddies in the flow. This makes the scaling of such a wall model independent of $Re_\tau$ over the range of $Re_\tau$ values of interest in external aerodynamics applications. This physics-based multiscale modeling approach that couples DNS and LES should be contrasted with other similar patch ideas whose size is linked to the LES grid size, resulting in an unfavorable scaling of the number of grid points required with respect to $Re_\tau$, being $Re_\tau^\alpha$ with $1 \leq \alpha \leq 2$  for wall-resolved LES resolutions \citep{Sandham2017}. Figure \ref{fig:Coupled model} illustrates all the components of the proposed coupling. The proposed self-similarity-based bridge between the two domains can also be viewed as an alternative to multi-block methods \citep{Pascarelli2000}.

The remainder of this study is structured as follows. Section 2 presents the outer flow LES formulation for a channel. Section 3 presents the formulation of the DNS near-wall patch wall model along with an attached-eddy-based coupling between the two simulation domains. Section 4 presents \textit{a-priori} results of applying the attached-eddy-based boundary conditions to the DNS near-wall patch. Finally, conclusions are drawn and future tasks are presented in Section 5.

\section{Outer flow formulation in channel flow}
Consider the continuity and the incompressible Navier-Stokes equations
\begin{equation}\label{Continuity}
    \frac{\partial u_i}{\partial x_i} = 0    
\end{equation}
\begin{equation}\label{NavierStokes}
    \frac{\partial u_i}{\partial t}+u_j\frac{\partial u_i}{\partial x_j}=-\frac{\partial p}{\partial x_i}+\nu\frac{\partial^2 u_i}{\partial x_j \partial x_j}\mathrm{,}
\end{equation}
where $u_i$ is the velocity component in the $i$th direction; $x_1,x_2, \mathrm{and}~x_3$ are the streamwise, wall-normal, and spanwise directions, respectively; $\nu$ is the kinematic viscosity; and $p$ is the kinematic pressure. Defining $\widehat{(.)}$ to be the filtering operation, filtering Eqs. \eqref{Continuity}-\eqref{NavierStokes} leads to the LES governing equations
\begin{equation}\label{ContinuityLES}
    \frac{\partial \widehat{u}_i}{\partial x_i} = 0    
\end{equation}
\begin{equation}\label{NavierStokesLES}
    \frac{\partial \widehat{u}_i}{\partial t}+\widehat{u}_j\frac{\partial \widehat{u}_i}{\partial x_j}=-\frac{\partial \widehat{p}}{\partial x_i}+\nu\frac{\partial^2 \widehat{u}_i}{\partial x_j \partial x_j}-\frac{\partial \tau_{ij}}{\partial x_j}\mathrm{,}
\end{equation}
where $\tau_{ij} = \widehat{u_i u_j}-\widehat{u}_i\widehat{u}_j$ is the SGS stress tensor. In the bulk of the channel, Eqs. \eqref{ContinuityLES}-\eqref{NavierStokesLES} are solved on an isotropic grid. The channel is driven by a constant mass flux such that the wall-shear stress is not fixed \textit{a-priori}. Since the LES equations are solved on a grid resolving only the large eddies in the logarithmic region of the flow, the appropriate boundary condition to apply at the wall for the tangential velocities is not the no-slip boundary condition. Instead, a boundary condition is placed on the wall-shear stress, $\tau_w^{LES}$. Most commonly, a no-penetration boundary condition at the wall for the filtered velocity is applied, forcing the wall-shear stress to be carried by resolved shear and SGS stress contributions. For eddy-viscosity-type SGS models, this makes the boundary condition a Neumann-type boundary condition for the tangential velocities at the wall. A wall model, which is usually based on RANS formulations, is needed to compute $\tau_w^{LES}$. Instead, we extract $\tau_w^{LES}$ from the DNS near-wall patch, which acts as the wall model. Due to the homogeneity of the channel in the wall-parallel directions, the same wall-shear stress is applied at every wall grid point in the LES, and a single DNS near-wall patch is simulated. 


\section{A near-wall patch wall model}
A near-wall patch of DNS resolution is employed to model the near-wall dynamics. The main aim of this near-wall patch is to predict the wall-shear stress and one-point statistics such as the mean velocity profiles and turbulence intensities. Furthermore, we aim to accurately predict the velocity spectra. To make this model tractable, the cost of the near-wall DNS patch should scale more favorably with $Re_\tau$ than the cost of the LES to eventually make its cost marginal. To satisfy this constraint, the size of the near-wall patch must be independent of the LES grid, and it is chosen to be fixed in inner units such that the physical size of the DNS patch shrinks as $Re_\tau$ increases. The near-wall patch is designed to capture the near-wall self-sustaining cycle, along with a portion of the self-similar hierarchy of eddies such that a coupling with the LES flow field that utilizes information from both flow domains is formulated. Figure \ref{fig:Coupled model} illustrates the coupling between the DNS near-wall patch and the LES flow field. 

\subsection{A near-wall patch in inner units}
Consider a region near the wall that is homogeneous in both the streamwise and the spanwise directions. The wall-parallel averaged wall-shear stress $\overline{\tau_w}^{1,3}$ can be used to find a time-varying  friction velocity $u_\tau(t) = (\overline{\tau_w}^{1,3}/\rho)^{1/2}$, where $\overline{(.)}^{1,3}$ denotes averaging along the homogeneous directions, and $\rho$ is the fluid density. Rescaling Eqs. \eqref{Continuity}-\eqref{NavierStokes} with this time-dependent friction velocity, and its corresponding time-dependent friction length scale $\delta_v(t) = \nu/u_\tau(t)$, leads to
\begin{equation}\label{ContinuityPlus}
    \frac{\partial u^+_i}{\partial x^+_i} = 0 \mathrm{,}    
\end{equation}
\begin{equation}\label{NavierStokesPlus}
    \frac{\partial u^+_i}{\partial t^+}+u^+_j\frac{\partial u^+_i}{\partial x^+_j}=-\frac{\partial p^+}{\partial x^+_i}+\frac{\partial^2 u^+_i}{\partial x^+_j \partial x^+_j} - u_i^+\frac{1}{u_\tau}\frac{d u_\tau}{d t^+}\mathrm{,}
\end{equation}
where $(.)^+$ indicates time-dependent plus units. The linear forcing term on the right-hand side of Eq. \eqref{NavierStokesPlus} resembles the linear forcing term found in forced homogeneous isotropic turbulence \citep{Lundgren2003}. It is used to satisfy the constraint that $\partial u_1^+/\partial x^+_2|_w = 1$ at all times, which follows directly from the definition of plus units, ensuring that this DNS near-wall patch is fixed in inner units but varies in size in outer units as the flow develops. Note that as the near-wall patch reaches statistical equilibrium, this linear forcing term vanishes on average. The friction velocity $u_\tau(t^+)$ can be found as 
\begin{equation}
    u_\tau(t^+) = u_\tau(0)\exp\bigg(\int_{0}^{t^+}A(t'^+)dt'^+\bigg) \mathrm{,}
\end{equation}
where
\begin{equation}
    A \equiv \frac{1}{u_\tau}\frac{d u_\tau}{dt^+} = -\frac{1}{U_b^+}\bigg[\frac{d U_b^+}{d t^+}+\frac{\overline{u_1^+u_2^+}^{1,3}}{\delta_P^+}\bigg|_{\delta_P^+}+\frac{dP^+}{dx_1^+}-\frac{1}{\delta_P^+}\bigg(\frac{\partial \overline{u^+}^{1,3}}{\partial x_2^+}\bigg|_{\delta_P^+}-1\bigg)\bigg] \mathrm{,}
\end{equation}
$U_b^+$ is the bulk velocity of the near-wall patch in inner units, $\delta_P^+$ is the wall-normal extent of the DNS near-wall patch in inner units, which is a parameter to be defined, $dP^+/dx_1^+ = d\hat{P}^+/dx_1^+$ is the time-varying mean pressure gradient driving the constant mass flux LES channel, and $u_\tau(0)$ is the friction velocity of the initial condition of the near-wall patch. Given this formulation, $\tau_w^{LES}$ can be defined as $\tau_w^{LES} = \rho u_\tau^2$, completing the formulation for the LES domain. 

The near-wall patch is homogeneous in the $x_1$ and $x_3$ directions, and periodic boundary conditions are applied to all velocity components. In the wall-normal direction, the no-slip and no-penetration boundary conditions are applied at the bottom wall. However, the correct boundary condition to apply at the upper boundary is ambiguous, as it is placed a distance $\delta_P(t)$ away from the wall, within a highly turbulent region. Specifying this top boundary condition for the three velocity components closes the system. In the following subsection, a physics-based approach utilizing the self-similar nature of the logarithmic region with distance from the wall is pursued to obtain suitable top boundary conditions for the DNS patch. The DNS near-wall patch of \cite{Carney2020} was also fixed in inner units, but instead of being predictive of the wall-shear stress, the wall-shear stress was a parameter defining the turbulent environment. 

\subsection{Attached-eddy-based coupling between the LES and the near-wall patch}

To specify the boundary conditions of the three velocity components at the top of the near-wall patch, we separately specify the mean and fluctuating components of each of the three velocity components 
\begin{equation}
    u_i|_{\delta_P} = \overline{u}_i|_{\delta_P} + u_i'|_{\delta_P}\mathrm{.}
\end{equation}
To specify the mean component of the velocities $\overline{u}_i$, a logarithmic mean velocity profile is assumed to exist between the location in the outer LES flow $x_2^{AE}$, from where statistical information is extracted and to be defined in the following paragraphs, and the top of the DNS patch $\delta_P$
\begin{equation}\label{LogMean}
    \frac{\overline{u_i}^{1,3}(x_2^{AE})-\overline{u_i}^{1,3}(\delta_P)}{u_\tau} = -\frac{1}{\kappa}\log\bigg(\frac{\delta_P}{x_2^{AE}}\bigg)+B^* ~~~ \mathrm{for} ~~~  i = 1,3,
\end{equation}
where $\kappa$ is the von K\'arm\'an constant and $B^*$ is the intercept of the log-law. The parameters $\kappa$ and $B^*$ can be found dynamically from the outer LES velocity using a least-squares method centered around $x_2^{AE}$. The mean of the wall-normal velocity has to be zero due to the symmetry of the flow. 

To fully define the fluctuating component of each of the velocities at the top boundary, a magnitude and a spatial structure need to be specified
\begin{equation}\label{FluctuatingBC}
    u_i'|_{\delta_P} = \alpha_i u_i^*|_{\delta_P}\mathrm{,}
\end{equation}
where $\alpha_i$ and $u_i^*|_{\delta_P}$ are the magnitude and spatial structure of the $i^{th}$ fluctuating velocity component at the top boundary, respectively. To specify each of them, we appeal to the self-similar structure of the flow. To specify the magnitude, the statistical attributes of a geometrically self-similar flow are leveraged. For any wall-bounded flow, the wall-normal behavior of the turbulent intensities can generally depend on the distance from the wall, the Reynolds number, the mean pressure gradients, and any other environment defining variables
\begin{equation}\label{GeneralAEScaling}
\begin{gathered}
    \frac{\overline{u_1'^2}}{u_\tau^2} = f_1\bigg(\frac{x_2}{\delta},Re_\tau,\frac{dP}{dx_1},\frac{dP}{dx_3},\ldots\bigg) \mathrm{,} \\
    \frac{\overline{u_3'^2}}{u_\tau^2} = f_3\bigg(\frac{x_2}{\delta},Re_\tau,\frac{dP}{dx_1},\frac{dP}{dx_3},\ldots\bigg)\mathrm{,} \\
    \frac{\overline{u_2'^2}}{u_\tau^2} = f_2\bigg(\frac{x_2}{\delta},Re_\tau,\frac{dP}{dx_1},\frac{dP}{dx_3},\ldots\bigg)\mathrm{,}
\end{gathered}
\end{equation}
where $\delta$ is the channel half-height or boundary layer thickness, and $\overline{(.)}$ denotes averaging along the homogeneous directions and time. At sufficiently high $Re_\tau$, and when there is a constant stress layer in the flow, the logarithmic layer is hypothesized to be mostly composed of wall-attached eddies that are geometrically self-similar with distance from the wall \citep{Townsend1976,Perry1982,Marusic2019}. These wall-attached eddies carry the bulk of the turbulent kinetic energy and momentum of the flow. In this preliminary study, we focus on tackling the coupling problem between the DNS near-wall patch and the LES domain in this limit of high-$Re_\tau$ flows. Under these conditions, the turbulent intensity profiles in Eq. \eqref{GeneralAEScaling} reduce to $Re_\tau$-independent, logarithmic forms
\begin{equation}\label{AEScaling}
\begin{gathered}
    \frac{\overline{u_1'^2}}{u_\tau^2} = B_1 - A_1 \log\bigg(\frac{x_2}{\delta}\bigg) \mathrm{,} \\
    \frac{\overline{u_3'^2}}{u_\tau^2} = B_2 - A_2 \log\bigg(\frac{x_2}{\delta}\bigg) \mathrm{,} \\
    \frac{\overline{u_2'^2}}{u_\tau^2} = B_3,
\end{gathered}
\end{equation}
where $A_1$, $A_2$, $B_1$, $B_2$, and $B_3$ are coefficients that depend on the flow.

Since wall-attached eddies are self-similar with distance from the wall, and they represent the largest fully confined eddy below a given wall-normal height, a carefully chosen LES resolution can resolve them. \cite{Lozano-Duran2019} showed that to resolve $90\%$ of the turbulent kinetic energy at some $x_2$ while recovering two-dimensional energy spectra resembling those of DNS except at the smallest scales, the grid resolution needs to satisfy
\begin{equation}\label{SpectraEstimates}
    \frac{2\Delta_1^{min}}{x_2}\approx 0.15\mathrm{,} ~~~  \frac{2\Delta_3^{min}}{x_2}\approx 0.15\mathrm{,}
\end{equation}
where $\Delta_1$ and $\Delta_3$ are the grid sizes in the streamwise and spanwise directions, respectively. Given that the wall-attached eddy scaling is valid at high $Re_\tau$ in the range of $2.6 Re_\tau^{1/2} \leq x_2^+ \leq 0.15 Re_\tau$, as shown by \cite{Klewicki2009}, then to resolve the scales at some $x_2$ in this region, assuming an isotropic grid as is usually employed in WMLES, requires approximately $90$ points in the wall-normal direction. This range of the validity of the attached-eddy scaling places the constraint on the value of $\delta_P^+ \geq 2.6 Re_\tau^{1/2}$. For the $Re_\tau$ values of interest for external aerodynamics, $Re_\tau \sim O(10^4-10^5)$ \citep{Smits2013}, choosing $\delta_P^+ \sim 1000$ makes the cost of the DNS patch independent of $Re_\tau$.  

By resolving these wall-attached eddies in the LES flow field around $x_2^{AE}$ in the range $2.6 Re_\tau^{1/2} \leq x_2^{AE} \leq 0.15 Re_\tau$, the coefficients $u_\tau^2A_1$, $u_\tau^2A_2$, $u_\tau^2B_1$, $u_\tau^2B_2$, and $u_\tau^2B_3$ can be extracted dynamically from the LES solution and Eq. \eqref{AEScaling} using a least-squares method centered around $x_2^{AE}$, where $u_\tau$ is computed using the DNS near-wall patch. These predicted coefficients provide the first piece of information, i.e. the statistics required to specify the magnitude of the fluctuating component of the velocity boundary conditions at the top of the DNS near-wall patch. To determine $\alpha_i$, any statistical quantities that can be predicted using Eq. \eqref{AEScaling} such as the turbulent intensity or the slope of the turbulent intensity of each velocity component can be utilized. In this study, we chose to hold the slopes of the turbulent intensities constant for the two wall-parallel components of the fluctuating velocities
\begin{equation}\label{SlopeBC}
    \frac{\partial\overline{u_i'^2}}{\partial x_2}\bigg|_{\delta_P} = \gamma_i ~~~ \mathrm{for} ~~~  i = 1,3,
\end{equation}
making $\alpha_i = \alpha_i(\gamma_i)$. Thus, Neumann boundary conditions are enforced for the slope of the mean turbulent intensities in the wall-parallel directions. The magnitude of the wall-normal fluctuating velocity cannot be specified using this statistical constraint as explained in the following paragraph.

The aim is to find the boundary condition that least perturbs the flow, allowing for a smooth continuation in the internal solution up to the boundary. To do so, the spatial structures of the fluctuating velocities in Eq. \eqref{FluctuatingBC} are chosen such that the interior of the domain is connected with the top boundary using wall-normal geometric self-similarity. Given that $\delta_P^+ \geq 2.6 Re_\tau^{1/2}$, the self-similar hierarchy of eddies extends beneath the top boundary of the DNS domain. Therefore, copying internal planes of fluctuating velocities and scaling them upward in size, could serve as potential boundary conditions. For simplicity, we copy the first internal plane below the top boundary for each velocity component
\begin{equation}\label{InternalCopyBC}
     u_i^*|_{\delta_P} \propto u_i'|_{\delta_P-\Delta_{x_2}},
\end{equation}
where $\Delta_{x_2}$ is the grid spacing at the top of the domain in the wall-normal direction, and the proportionality symbol indicates that only the spatial structure is extracted by rescaling the variance of the copied internal plane to unity. This choice of boundary condition assumes that the DNS grid is sufficiently resolved in the wall-normal direction such that wall-parallel scaling of the size of the structures with distance from the wall is unnecessary. However, if an interior plane farther away from the top boundary is copied, the wall-parallel scaling of the structure of the flow will become necessary to maintain the geometric self-similarity that is the basis for this boundary condition. The combination of Eqs. \eqref{FluctuatingBC},\eqref{AEScaling}, and \eqref{InternalCopyBC} imply that a zero Neumann boundary condition should be applied to the instantaneous wall-normal velocity component, which is not physical. In this preliminary study, we explore the use of a Robin-type boundary condition for the wall-normal velocity component
\begin{equation}\label{SlipLength}
    u_2 = \ell \frac{\partial u_2}{\partial x_2}\mathrm{,}
\end{equation}
where the slip length $\ell$ is a parameter to be specified. 

Given this two-sided enforcement of the wall-normal geometric self-similarity as a boundary condition, the system for the DNS near-wall patch is closed, and the coupling between the LES domain and the DNS domain is established in a dynamic fashion, relying on no \textit{a-priori} picked coefficients, except for the choice of $\ell$. However, in principle, a purely parameter-free coupled system can be formulated which incorporates a dynamic boundary condition for the wall-normal velocity. Possible methods include supplementing the attached-eddy-based scalings in the wall-parallel directions with a statistical constraint on the wall-normal turbulent intensity to allow the application of Eqs. \eqref{FluctuatingBC} and \eqref{InternalCopyBC} directly, or the internal rescaling of planes further inside the DNS patch domain. In future versions of this work, these dynamic boundary conditions for the wall-normal velocity will be investigated. In the next section, we present \textit{a-priori} results of applying the boundary conditions to a truncated channel flow.

An alternative type of boundary condition that was tested at the initial stages of the investigation was a one-way enforcement of the wall-normal geometric self-similarity, where both the magnitude and the structure of the fluctuating velocities in Eq. \eqref{FluctuatingBC} were extracted from the outer LES flow. An entire plane of fluctuating velocities from the LES solution at $x_2^{AE}$ was extracted, scaled linearly with distance from the wall in the wall-parallel directions, scaled in amplitude using Eq. \eqref{AEScaling}, evaluated at $x_2 = \delta_P^+$, and translated in the wall-parallel directions by the logarithmic mean velocities in Eq. \eqref{LogMean} to account for the difference in the advection velocity of the fluctuations at different wall-normal heights. These modified velocity planes were then applied as Dirichlet boundary conditions to the top of the DNS near-wall patch. However, due to the disconnect between these planes and the internal flow field of the DNS near-wall patch, spurious pressure fluctuations were introduced that modified the mean velocity profile at distances proportional to $\delta_P$ \citep{Jimenez1998,Mizuno2013}. For this reason, this boundary condition was not pursued any further. 


\section{\textit{A-priori} testing of the attached-eddy-based boundary conditions}
\begin{figure}
    \centering
    \includegraphics[width=\textwidth]{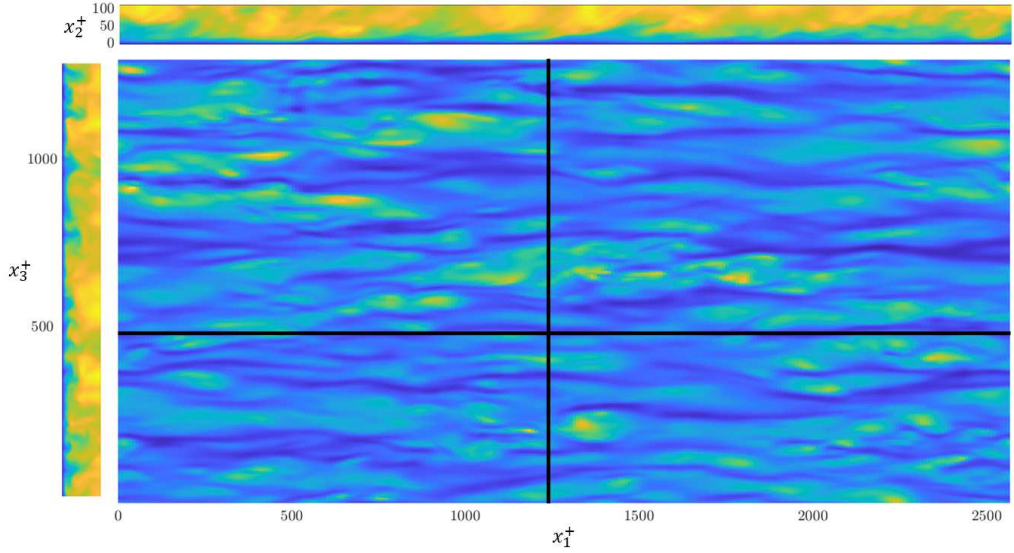}
    \caption{Instantaneous streamwise velocity in a the DNS near-wall patch simulation at a nominal $Re_\tau = 395$. Low velocities are indicated by darker colors and higher velocities with lighter colors. The flow in the $x_1-x_3$ plane shows the near-wall streaky structure of the flow, the flow in the $x_1-x_2$ plane shows ejections and sweeps, and the flow in the $x_2-x_3$ plane shows the low-speed streaks due to the lift-up effect of streamwise-aligned vortices. The two solid lines indicate the locations at which the cross-sections are taken.}
    \label{fig:DNSFlowStructurePatch}
\end{figure}

\begin{figure}
     \centering
     \begin{subfigure}[b]{0.49\textwidth}
         \centering
         \includegraphics[width=\textwidth]{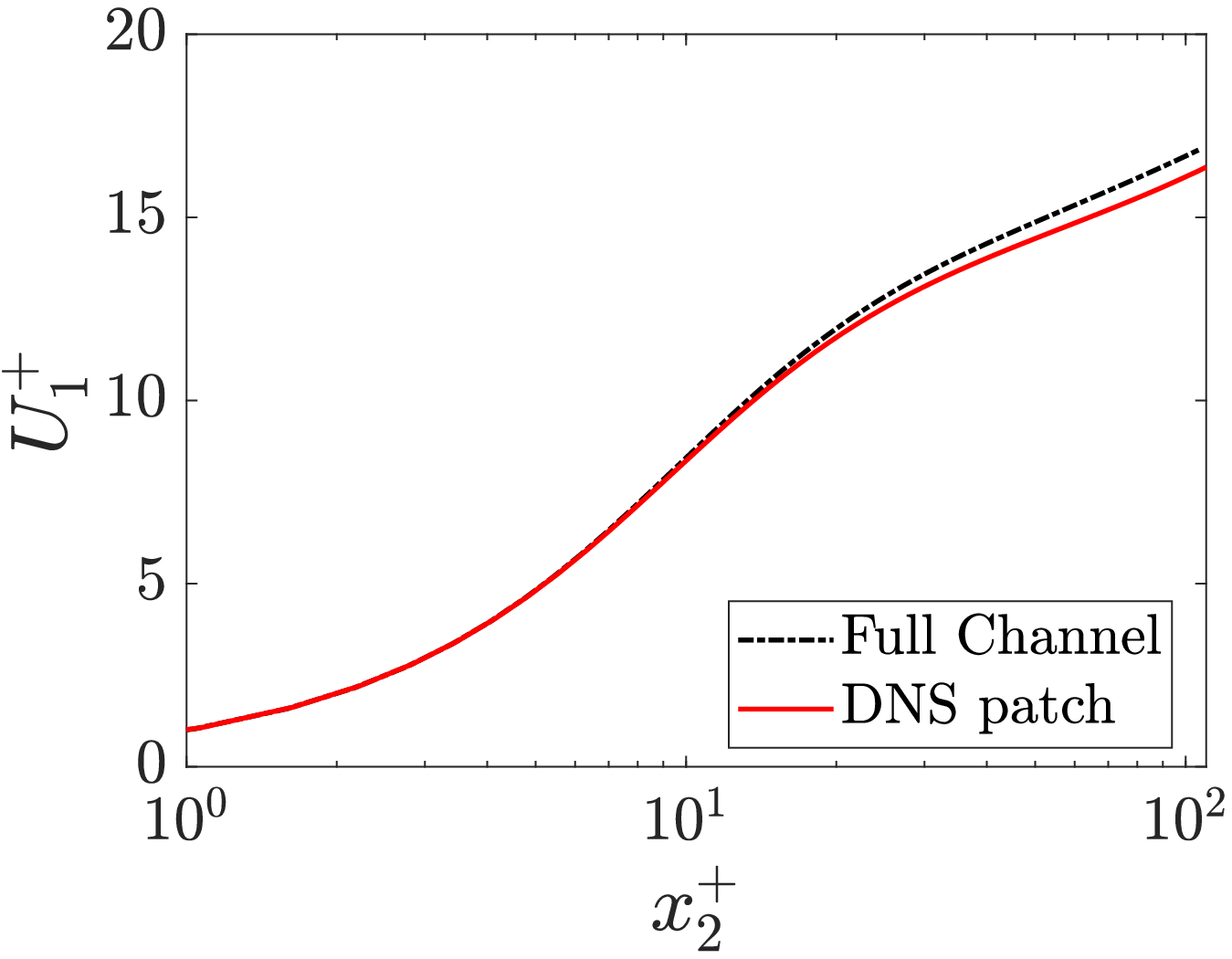}
         \caption{}
         \label{fig:MeanVel}
     \end{subfigure}
     \hfill
     \begin{subfigure}[b]{0.49\textwidth}
         \centering
         \includegraphics[width=\textwidth]{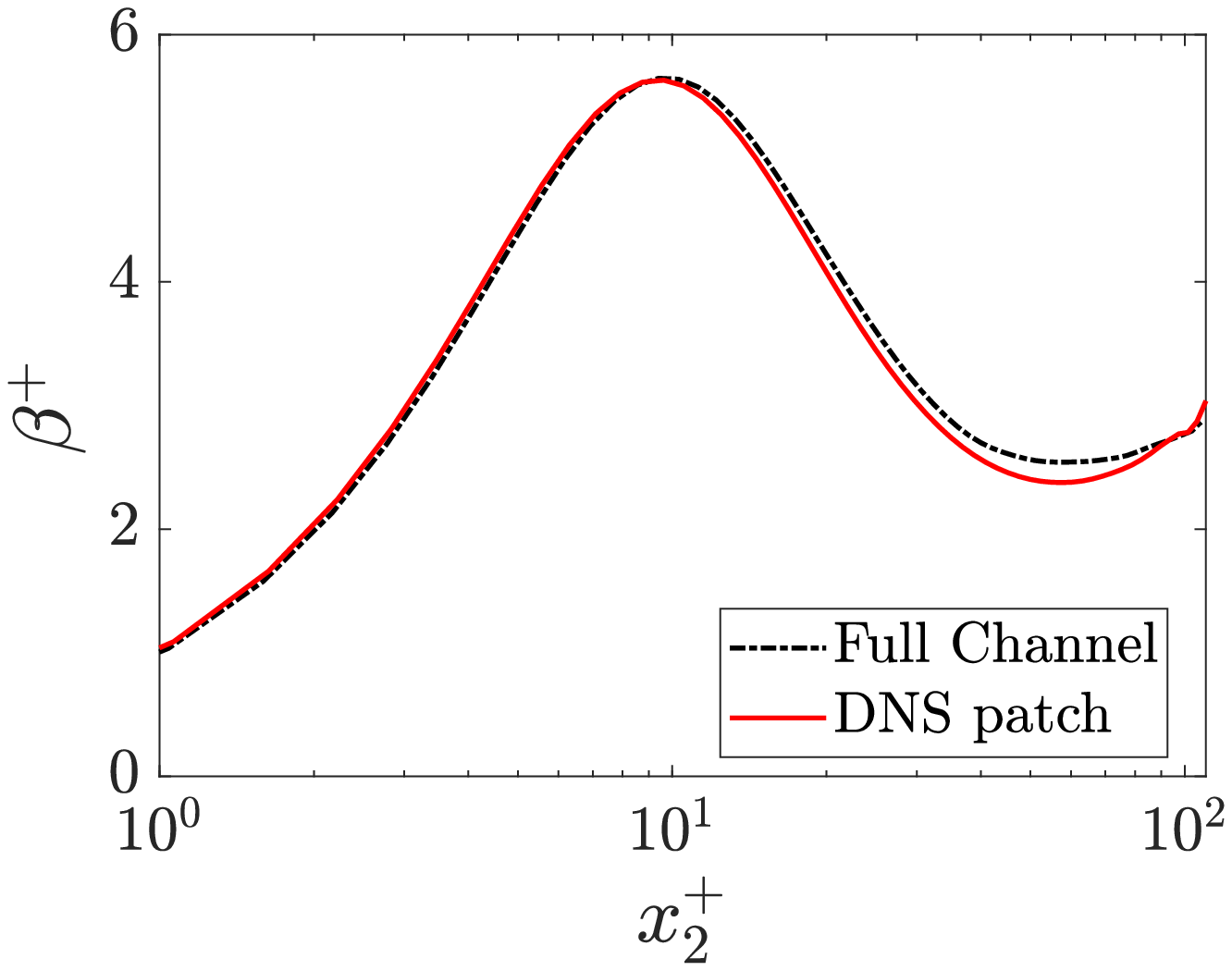}
         \caption{}
         \label{fig:IndicatorFunction}
     \end{subfigure}
        \caption{First-order one-point statistics of the full channel and the surrogate DNS near-wall patch. (a) Streamwise mean velocity profile. (b) Indicator function $\beta^+$. The indicator function $\beta^+ = x_2^+ \partial U_1^+ / \partial x_2^+$ is used to test for logarithmic regions in the mean velocity profile $U_1^+(x_2^+)$.}
        \label{fig:1st-order stats}
\end{figure}

To test the ability of the boundary conditions in Eqs. \eqref{InternalCopyBC}-\eqref{LogMean} applied within the logarithmic region of the flow in recovering realistic turbulent statistics inside the DNS near-wall patch, an \textit{a-priori} test is conducted at a low $Re_\tau = 395$. DNS of a full channel flow at $Re_\tau = 395$ with $(L_{x_1},L_{x_2},L_{x_3}) = (2\pi,2,\pi)$ is conducted with uniform grid resolutions in the wall-parallel directions of $(\Delta_{x_1}^+,\Delta_{x_3}^+) = (9.74, 4.87)$ and with a minimum and maximum grid resolution in the wall-normal direction of $\Delta_{x_2,min}^+ = 0.51$ and $\Delta_{x_2,max}^+ = 4.21$, respectively. Once statistical equilibrium is reached, the channel is truncated at $x_2^+ \approx 110$, a position close to the minimum of the indicator function defined in Figure \ref{fig:1st-order stats}, an incipient logarithmic region of sorts, and the values of the mean velocity in the wall-parallel directions as well as the slopes of the turbulent intensities in Eq. \eqref{SlopeBC} are extracted from the full channel and applied as boundary conditions to the truncated domain. This truncated channel, a surrogate for the DNS near-wall patch, is run for $15$ eddy turnover times to statistical equilibrium using the truncated channel flow as an initial condition. Given that the slip length in Eq. \eqref{SlipLength} is a free parameter yet to be specified, a sweep was conducted with $\ell \in \{-0.1,-0.2,-0.4,-0.8,-1.6,-3.2\}$ and it was found that the first-order quantities of interest such as the mean velocity profile and $u_\tau$ do not significantly change. As such, the results presented are for $\ell = -0.4$. Furthermore, all results are presented in $(.)^+$ units normalized by the respective friction velocities of each simulation. Finally, note that because the initial condition was statistically close to the final equilibrium solution, the linear forcing term in Eq. \eqref{NavierStokesPlus} was not included as it should be small. Figure \ref{fig:DNSFlowStructurePatch} show instantaneous snapshots of the streamwise velocity in wall-parallel, cross-stream, and wall-normal streamwise-aligned planes. 

For the cases tested, $u_\tau$ of the DNS patch was always predicted within $5\%$ of that of the full channel, showing initial success of the top boundary conditions. Figure \ref{fig:1st-order stats}(a) shows the streamwise mean velocity profile from the full channel truncated to $x_2^+ = 110$ as well as the mean velocity profile of the surrogate DNS near-wall patch. It is evident that the boundary conditions do not introduce significant artifacts, such as the emergence of an internal adaptation layer at the top of the domain, and that the mean velocity profile of the DNS patch tracks that of the full channel across the entirety of the domain. This is verified in Figure \ref{fig:1st-order stats}(b), which compares the logarithmic indicator function of the DNS patch with that of the full channel. The indicator function of the DNS patch tracks that of the full channel across the entirety of the domain and captures the minimum in the indicator function within the correct wall-normal region away from the wall. This minimum is underpredicted by about $7\%$. Note that there is no true logarithmic region in the mean velocity profile at this low $Re_\tau$. However, as the magnitude of $\ell$ was increased, an artifical logarithmic region with an overpredicted von K\'arm\'an constant, where the indicator function attained a constant value with wall-normal distance, developed towards the edge of the domain. 

\begin{figure}
    \centering
    \includegraphics[width=\textwidth]{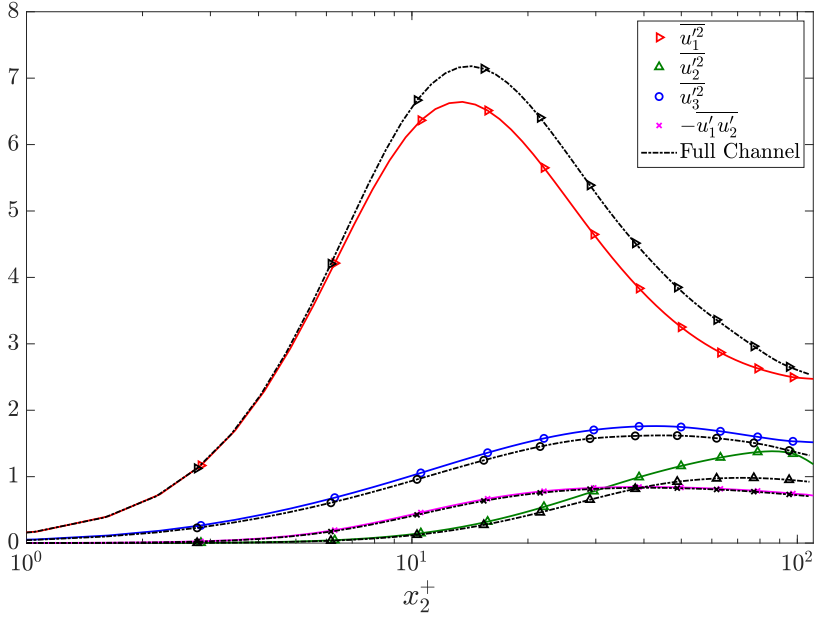}
    \caption{A comparison of the wall-normal profiles of the four nonzero components of the Reynolds stress tensor in the full channel with those in the DNS near-wall patch. The solid lines represent the DNS patch and the dash-dotted lines with matching symbols represent the full channel simulation at $Re_\tau = 395$.}
    \label{fig:Turbulent Intensities}
\end{figure}

Figure \ref{fig:Turbulent Intensities} shows the turbulent intensities predicted by the DNS patch. Both the streamwise and the spanwise turbulent intensities are predicted with reasonable accuracy across the entire wall-normal extent of the DNS patch domain. However, there is a large overprediction of the wall-normal turbulent intensity near the top boundary that extends towards the wall. This overprediction in the wall-normal turbulent intensity could be tied to the underprediction of the streamwise intensity through pressure-strain correlation and is most likely related to the choice of the Robin-type boundary condition for the wall-normal velocity, as it was the quantity most sensitive to the value of $\ell$. Furthermore, even though no explicit constraints were placed on the Reynolds shear stress, it is predicted accurately across the entire wall-normal distance. This high level of accuracy in the prediction of the Reynolds shear stress is explained through the mean momentum balance, which ties the Reynolds shear stresses to the mean velocity profile, which was predicted with accurately. The emergence of the artifical logarithmic region with the increased magnitude of $\ell$ did not affect the Reynolds stress wall-normal profile significantly. Note that the lowest error in the prediction of $\kappa$ and $u_\tau$ was achieved when the choice of $\ell$ made the wall-normal turbulent intensity closest to its correct value at the boundary. This suggests that the value of $\ell$ could be specified by enforcing the value of the wall-normal turbulent intensity at the boundary, which can be extracted from the wall-normal turbulent intensity from the LES flow at $x_2^{AE}$ in a fully coupled calculation.  

\begin{figure}
    \centering
    \makebox[\textwidth][c]{\includegraphics[width=1.15\textwidth]{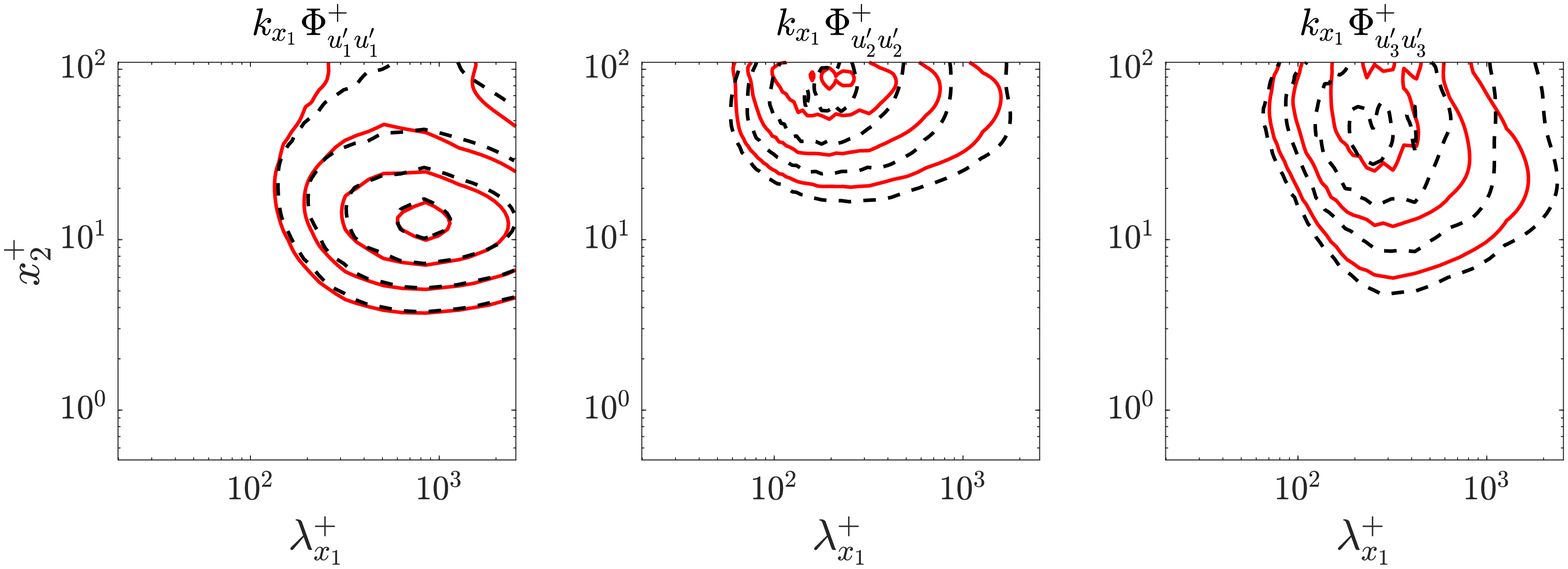}}
    \caption{Premultiplied one-dimensional energy spectra as a function of the wall-normal distance and the streamwise wavelength. The solid lines represent the DNS near-wall patch and the dashed lines represent the full channel simulation. The contour levels are $\{0.3,0.5,0.75,0.95\}\times\mathrm{max}(k_{x_1}\Phi_{u'_iu'_i}^+)$.}
    \label{fig:Pre-multiplied Streamwise Spectra}
\end{figure}

\begin{figure}
    \centering
    \makebox[\textwidth][c]{\includegraphics[width=1.15\textwidth]{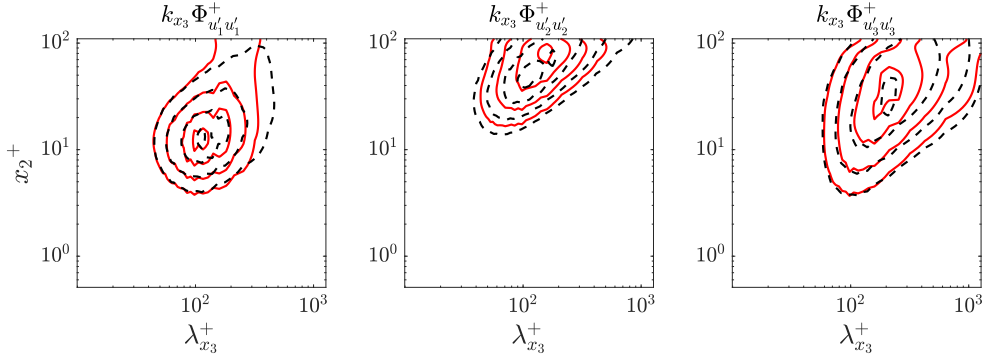}}
    \caption{Premultiplied one-dimensional energy spectra as a function of the wall-normal distance and the spanwise wavelength. The solid lines represent the DNS near-wall patch and the dashed lines represent the full channel simulation. The contour levels are $\{0.3,0.5,0.75,0.95\}\times\mathrm{max}(k_{x_3}\Phi_{u'_iu'_i}^+)$.}
    \label{fig:Pre-multiplied Spanwise Spectra}
\end{figure}

To compare the structure of the flow between the full channel and the surrogate DNS near-wall patch, contours of premultiplied one-dimensional energy spectra in both the streamwise and the spanwise directions are considered. Figures \ref{fig:Pre-multiplied Streamwise Spectra} and \ref{fig:Pre-multiplied Spanwise Spectra} show that the near-wall peak in the streamwise spectra of the streamwise turbulent intensity is captured accurately, and that the considered contour levels match approximately all the way to the top boundary of the patch. This is also approximately seen in the spanwise spectra of the streamwise turbulent intensity with any appreciable difference only observed toward the very top of the domain. Qualitatively, both the streamwise and the spanwise one-dimensional spectra of both the wall-normal and the spanwise turbulent intensities in the DNS patch match the full channel simulation, with two notable exceptions. First, the peak in the streamwise spectrum of the spanwise turbulent intensity is closer to the wall in the full channel than it is in the DNS patch, where it occurs at the boundary. Second, the same behavior can be observed in the spanwise spectrum of the wall-normal turbulent intensity, where the peak is closer to the top of the domain in the DNS patch than it is in the full channel simulation. 

\section{Conclusions and future work}

The current landscape of RANS-based wall models does not make use of any structural information of the flow. In this preliminary study, we examine the feasibility of a structure-based wall model that aims to capture the near-wall self-sustaining cycle and a portion of the wall-normal self-similar hierarchy of eddies, such that a coupling is possible with the largest eddies in this hierarchy resolved by the LES grid. A framework for coupling a patch of DNS resolution fixed in inner units to the outer LES flow is presented. Dynamically computed statistical quantities from the LES are applied as constraints to the self-similarity-based top boundary condition of this DNS patch. \textit{A-priori} testing of this boundary condition was conducted. It is found that copying internal planes from within the incipient logarithmic region of the flow does not introduce significant artifacts near the top boundary and recovers quantitatively accurate first- and second-order one-point statistics. The predicted friction velocity $u_\tau$ was within $5\%$ of the full channel flow, and the mean velocity profile, streamwise turbulent intensity, and the Reynolds shear stress were predicted accurately across the entire domain. Two-point statistics in the form of the streamwise and spanwise premultiplied energy spectra for all three velocity components compared quantitatively well with the full channel simulation and showed that the structure of the flow was also realistic up to the top of the domain.  The \textit{a-priori} tests were conducted at the relatively low $Re_\tau = 395$. Since the validity of the assumptions upon which the boundary conditions are based increases as $Re_\tau$ increases, the \textit{a-priori} tests are considered to be a stringent test of the proposed boundary conditions. It is hypothesized that the accuracy of the predictions would increase with $Re_\tau$.

Even though a general framework was presented, some assumptions that were made require refinement. For example, the attached-eddy-based form of the turbulent intensities is valid for the dynamically computed coefficients only at extreme values of $Re_\tau$. To extend this coupling to moderate values of $Re_\tau$ the form of the turbulent intensities can be enhanced by accounting for the existence of other types of eddies, such as wall-detached self-similar eddies \citep{Marusic2019}. Furthermore, enforcing other statistical quantities as constraints on the top boundary of the DNS patch, which would allow for the removal of the ad hoc choice of a Robin-type boundary condition for the wall-normal velocity, needs to be explored. Once all these refinements are made, a fully coupled calculation between an outer-flow LES simulation and the DNS near-wall patch will be conducted in both equilibrium and nonequilibrium settings.

\section*{Acknowledgments} 

This study was funded by the Stanford Engineering Graduate Fellowship and by NASA grant \#NNX15AU93A.

\bibliographystyle{ctr}
 







\bibliographystyle{ctr}


\end{document}